\documentclass[11pt]{article}
\usepackage{amsmath}
\usepackage{amssymb}
\usepackage{graphicx}
\def\xxx#1 {{\sf hep-th/#1} }

\def\D{\Delta}

\def\a{\alpha}
\def\b{\beta}
\def\g{\gamma}
\def\d{\delta}
\def\e{\varepsilon}
\def\m{\mu}

\def\s{\sigma}

\def\t{\tau}

\def\k{\kappa}

\def\mc{\mathcal}

\def\la{\langle}
\def\ra{\rangle}
\def\dag{\dagger}
\def\wt{\widetilde}
\def\da{\dot{\a}}
\def\db{\dot{\b}}
\def\dg{\dot{\g}}

\def\ds{\dot{\s}}
\numberwithin{equation}{section} 
\parskip=\medskipamount

\begin{document}
\thispagestyle{empty}
\addtocounter{page}{-1}
\def\thefootnote{\fnsymbol{footnote}}
\begin{flushright}
  hep-th/0407098 \\
  AEI-2004-052 \\
  UW/PT-04-08
\end{flushright}

\vskip 0.5cm

\begin{center}\LARGE
{\bf New aspects of the BMN correspondence \\ beyond the planar limit}
\end{center}

\vskip 1.0cm

\begin{center}
{\large Petra Gutjahr\footnote{E-mail address: {\tt petra.gutjahr@aei.mpg.de}} 
and Ari Pankiewicz\footnote{E-mail address: {\tt ari@phys.washington.edu}}}

\vskip 0.5cm

{\it $^*$ Max-Planck-Institut f\"ur Gravitationsphysik, Albert-Einstein
Institut \\ Am M\"uhlenberg 1, D-14476 Golm \rm GERMANY}

\vskip 0.5cm

{\it $^{\dag}$ Department of Physics, University of Washington \\ P.O. Box 351560, Seattle WA 98195 \rm USA}

\end{center}

\vskip 1.0cm

\begin{center}
July 2004
\end{center}

\vskip 1.0cm

\begin{abstract}
\noindent
Motivated by recent disagreements in the context of AdS/CFT, we study the non-planar sector of the BMN correspondence.
In particular, we reconsider the energy shift of states with two stringy excitations in light-cone string field theory and explicitly determine 
its complete perturbative contribution from the impurity-conserving channel. Surprisingly, our result neither agrees with earlier leading order 
computations, nor reproduces the gauge theory prediction. More than that, it features half-integer powers of the effective gauge coupling $\lambda'$ representing a 
qualitative difference to gauge theory. Based on supersymmetry we argue that the above truncation is not suited for conclusive tests of the BMN duality.    
\end{abstract}

\vfill

\setcounter{footnote}{0}
\def\thefootnote{\arabic{footnote}}
\newpage

\renewcommand{\theequation}{\thesection.\arabic{equation}}


\section{Introduction}\label{section1}
In the last two years a lot of progress has been made in understanding the ideas of the AdS/CFT correspondence. The starting point of a long chain of 
developments marks the BMN duality~\cite{Berenstein:2002jq}, which connects ${\cal N}=4$ $U(N)$ super Yang-Mills (SYM) in the sector of operators with large $U(1)_R$ charge 
$J\propto \sqrt{N} \to \infty$ and type IIB string theory on the maximally supersymmetric plane-wave background~\cite{Blau:2001ne}. In contrast to $AdS_5 \times S^5$, 
here the worldsheet $\s$-model reduces to a free theory in the light-cone gauge and thus, can easily be quantized~\cite{Metsaev:2001bj}. Moreover, string interactions can 
be treated in the context of light-cone string field theory~\cite{Spradlin:2002ar,Pankiewicz:2003kj,Chu:2002eu}.\\
The parameters of the two sides of the duality are linked 
via~\cite{Berenstein:2002jq,Constable:2002hw}\footnote{Here $\mu$ denotes the curvature scale of the plane-wave 
geometry and $p^+$ is the light-cone momentum.} 
\begin{equation}
\frac{1}{(\mu\, \a'\, p^+)^2} = \frac{g_{\rm YM}^2 \,N}{J^2} \equiv \lambda' \qquad \text{and} \qquad 4\, \pi\, g_s\, (\mu\, \a'\, p^+)^2  = \frac{J^2}{N} \equiv g_2\,,
\end{equation}
and therefore free string theory corresponds to the planar ($g_2 =0$) sector, while string interactions are identified with non-planar, interacting gauge theory.  
This statement has been subject to a wide range of tests based on the key relation 
\begin{equation}
\frac{1}{\mu}\, E_{\rm l.c.} = \D - J\, ,
\end{equation}
where $E_{\rm l.c.}$ is the light-cone energy of string states and $\D$ denotes the conformal dimension of the dual SYM-operators.
In particular for the planar case, $E_{\rm l.c.}$ could explicitly be verified for so-called two-impurity operators (i.e. two stringy excitations) 
up to two loops~\cite{Berenstein:2002jq,Gross:2002su,Beisert:2003tq}; for all-loop arguments see~\cite{Santambrogio:2002sb,Beisert:2004hm}. 
The extension to the interacting part proved to be much more involved. 
On the field theory side, the anomalous dimension of two-impurity operators is known up to ${\cal O}(g_2^2 {\lambda'}^2)$~\cite{Beisert:2002bb,Beisert:2002ff,Beisert:2003tq}
\begin{equation}\label{energy}
\D - J = 2 + \frac{g_2^2}{4\,\pi^2}\Big[ \Big(\frac{1}{12} + \frac{35}{32\, n^2 \, \pi^2}\Big) \lambda' + \frac{1}{4}\Big(\frac{1}{12} + \frac{35}{32\, n^2 \, \pi^2}\Big)^2 {\lambda'}^2 
+ \cdots \Big] \,, 
\end{equation} 
whereas in light-cone SFT only the leading order energy shift has been computed~\cite{Roiban:2002xr}. Over and above, matrix elements of $H_{\rm l.c.}$ and the dilatation 
operator $D$ as well as decay widths of one-string states and single-trace operators have been successfully compared to leading order, see e.g.~\cite{Gross:2002mh}. 
For more details we refer to the reviews~\cite{Pankiewicz:2003pg} and references therein.\\  
It was subsequently realized, that the BMN correspondence can be generalized in essentially two directions: Curvature corrections to the plane-wave background 
can be taken into account~\cite{Callan:2003xr} having their counterpart in $1/J$ corrections of planar BMN gauge theory~\cite{Beisert:2002tn,Beisert:2003ys}. 
On the other hand, a completely new field was established by considering semiclassical string states in $AdS_5 \times S^5$~\cite{Gubser:2002tv} and 
the insight that the planar spectrum of SYM operators is governed by an integrable Hamiltonian (long-range) spin chain~\cite{Minahan:2002ve}, leading to a vast class of 
tests, see the review~\cite{Tseytlin:2003ii} and references therein.      
Both cases show perfect agreement up to two loops~\cite{Arutyunov:2003rg}, see also~\cite{Kazakov:2004qf}, but -- despite of this tremendous progress -- several open 
questions and puzzles gradually emerged at three loops:
In the latter approach, string and gauge theory continue to exhibit qualitatively similar structures but start to differ in detail~\cite{Serban:2004jf}. 
Somewhat more disturbing results have been found in~\cite{Callan:2003xr}: degeneracies present in the gauge theory and crucial for 
integrability~\cite{Beisert:2003tq,Beisert:2003ys,Klose:2003qc} are lifted in the near plane-wave background. \\ 
These recent developments raise the question whether disagreements do already occur in the BMN duality itself, namely in the non-planar, interacting sector. 
As was pointed out in~\cite{Spradlin:2003bw}, certain matrix elements of $H_{\rm l.c.}$ and $D$ seem to mismatch starting at $g_2\,{\lambda'}^2$. This however need not necessarily 
imply that physical quantities show disagreement as well and deserves a more careful investigation, which we initiate in the present paper.
Eventually one would like to compute independently the energy of states with two (stringy) excitations on the string theory side and hopefully reproduce \eqref{energy}. \\
Therefore, in section 2, we briefly introduce some well-known facts about the free theory and explicitly construct the supermultiplet for states with two stringy 
excitations. Especially we find that states consisting of two fermionic/bosonic oscillators in general mix with each other.\\
The evaluation of the energy shift $\sim{\cal O}(g_2^2)$ demands the knowledge of cubic and quartic terms in the Hamiltonian. We review (section 3) the 
formulation of the three-string vertex restricted by the superalgebra at order $g_2$ and comment on further constraints on the quartic interaction. In particular, we notice
that a term induced by the second order dynamical supercharges cannot a priori be excluded.\\
It is a known fact, that all members of a supermultiplet receive the same energy corrections. Furthermore (section 4), one can show by using the states in the 
supermultiplet, degenerate perturbation theory becomes redundant. Note, that both statements are only valid when including impurity-conserving and 
-non-conserving intermediate states.
Here, we calculate as a first step the complete perturbative (in $\mu^{-1}$) impurity-conserving contribution for one particular representation. Quite surprisingly our result 
disagrees with that given in the literature~\cite{Roiban:2002xr} and also fails to reproduce the gauge theory prediction \eqref{energy} at two loops. 
Above all as a qualitative difference to gauge theory the series features not only integer, but also half-integer powers of $\lambda'$. 
It seems to be apparent that this truncated analysis does not reveal the whole story.         
We conclude with a discussion.


\section{The free theory}\label{section2}

In this section, we introduce the free theory and its symmetries and analyze the underlying supermultiplet structure.
Type IIB string theory on the plane-wave space-time can be quantized in light-cone gauge~\cite{Metsaev:2001bj} resulting in the Hamiltonian 
\begin{equation}
H_2 = \frac{1}{\a}\sum_{n\in\mathbb{Z}}\omega_n N_n\,,
\end{equation}
where $\a\equiv \a'p^+$ is the light-cone momentum, the frequencies are given by $\omega_n = \sqrt{n^2+(\m\a)^2}$
and $N_n$ indicates the number operator
\begin{equation}
N_n = \a_n^{\dag\,i}\a_n^i+\a_n^{\dag\,i'}\a_n^{i'}
+\bigl(\b_n^{\dag}\bigr)^{\a_1\a_2}\bigl(\b_n\bigr)_{\a_1\a_2}+\bigl(\b_n^{\dag}\bigr)^{\da_1\da_2}\bigl(\b_n\bigr)_{\da_1\da_2}\,.
\end{equation}
Compared to~\cite{Spradlin:2002ar} we performed a redefinition of the oscillator basis to have the 
standard level-matching condition (cf. Appendix \ref{appA}), i.e. the operator $\sum_{n\neq 0}nN_n$ 
has to vanish on the space of physical states.
Here the bosonic oscillators $\a_n$ transform as $\bf [4,1]=[(2,2),(1,1)]$ and $\bf [1,4]=[(1,1),(2,2)]$ under the transverse 
$SO(4)\times SO(4) \simeq SU(2)^2\times SU(2)^2$ isometry of the plane-wave background, whereas the 
fermionic oscillators $\b_n$ give the representations $\bf [(2,1),(2,1)]$ and $\bf [(1,2),(1,2)]$. 
Both obey the standard (anti)-commutation relations.\\
Due to the effective harmonic oscillator potential of the background geometry, the theory possesses an essentially unique $SO(4)\times SO(4)$ singlet ground state 
$|\a\ra$ (labeled by its light-cone momentum since $P^+$ is a central element of the plane-wave superalgebra) defined by 
\begin{equation}
\a_n|\a\ra = 0\,,\quad \b_n|\a\ra = 0\,,\quad n\in\mathbb{Z}\,.
\end{equation}
Excited states are obtained by acting with the creation oscillators on $|\a\ra$ subject to the level-matching condition and can be organized into 
multiplets of the plane-wave superalgebra. Its bosonic generators are $H$, $P^+$, $P^I$, $J^{+I}$ ($I=1,\ldots,8$) and the angular momentum 
generators of the transverse $SO(4)\times SO(4)$ $J^{ij}$ and $J^{i'j'}$, while
the 32 supersymmetries are generated by $Q^+$, $\bar{Q}^+$ (both transforming as $\bf [(2,1),(2,1)]$ and $\bf [(1,2),(1,2)]$)
and $Q^-$, $\bar{Q}^-$ (transforming as $\bf [(2,1),(1,2)]$ and $\bf [(1,2),(2,1)]$). \\
On a general eigenstate of $H_2$ in a given irreducible representation of $SO(4)\times SO(4)$ the nontrivial action of generators is as follows: 
Certain combinations of $P^I$ and $J^{I+}$ add or remove a bosonic zero-mode excitation; this raises or lowers the energy of the state by $\m$ and is the 
discretized analog of giving a state transverse momentum. Note, that $P^I$ is not a quantum number in the plane-wave space-time since it does not commute with $H$.
Similarly $Q^+$ and $\bar{Q}^+$ add or remove a fermionic zero-mode excitation. 
This is why not the energy but $C \equiv \sum_{n\neq 0}\omega_nN_n$, which only counts the non-zero-mode (`stringy') excitations, is a Casimir of 
the superalgebra.
Finally, the most interesting generators are $Q^-$ and $\bar{Q}^-$; these do not change the number of excitations, commute with $H$ and, therefore transform states 
of the same energy but different $SO(4)\times SO(4)$ representations into each other. This action of the generators is schematically depicted in Fig. 1.
\begin{figure}
\begin{center}
\includegraphics{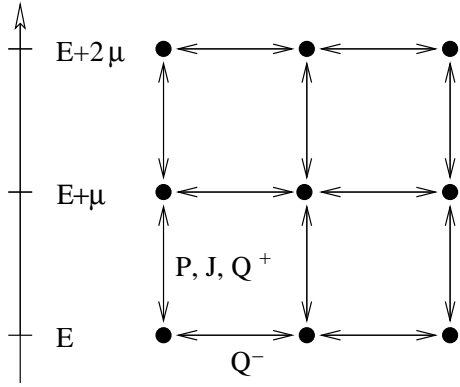} 
\end{center}
\caption{Action of the generators on the supermultiplet. Note, that the horizontal action of $Q^-$ terminates after at most eight units, while the vertical action
continues indefinitely.} 
\end{figure}
In the following we will call a multiplet containing states with $n$ stringy excitations a `$n$--impurity' multiplet.
The simplest example is the 'zero-impurity', i.e. supergravity multiplet~\cite{Metsaev:2002re}. In this 
case the highest weight state is the ground state $|\a\ra$ annihilated by all $Q^-$, $\bar{Q}^-$ and, therefore, 
this multiplet is short and protected against quantum corrections through string interactions.

\subsection{The two-impurity supermultiplet}

In contrast to the supergravity multiplet, where the state of lowest energy is unique, for the two-impurity multiplet 
the states of lowest energy $2\frac{\omega_n}{|\a|}$ are those with two stringy oscillators, schematically 
\begin{equation*}
\underline{\text{bosons}}:\qquad\a_n^{\dag}\a_{-n}^{\dag}|\a\ra\,,\quad \b_n^{\dag}\b_{-n}^{\dag}|\a\ra\,,\hspace{1cm}
\underline{\text{fermions}}:\qquad\a_n^{\dag}\b_{-n}^{\dag}|\a\ra\,,\quad \a_{-n}^{\dag}\b_n^{\dag}|\a\ra\,.
\end{equation*}
All of these 256 states are linked to each other by acting with half of the dynamical supercharges on a highest weight state which we will explicitly determine below, see 
also~\cite{Callan:2003xr} for the discussion in the case of the near plane-wave background.
States with two bosonic oscillators are decomposed into irreducible representations of $SO(4)\times SO(4)$, namely
\begin{align}
|{\bf [1,1]}\ra & = \frac{1}{2}\a_n^{\dag\,k}\a_{-n}^{\dag\,k}|\a\ra\,,\\ 
{|\bf[3^{\pm},1]\ra}^{[ij]} & = \frac{1}{2}\left(\a^{\dag\,i}_n\a^{\dag\,j}_{-n}-\a^{\dag\,j}_n\a^{\dag\,i}_{-n}\pm\e^{ijkl}\a^{\dag\,k}_n\a^{\dag\,l}_{-n}\right)|\a\ra\,,\\
{|\bf [4,4]\ra}^{ij'}_{\pm} & = \frac{1\pm\Omega}{\sqrt{2}}\a_n^{\dag\,i}\a_{-n}^{\dag\,j'}|\a\ra\,,\\
{|\bf [9,1]\ra}^{(ij)} & = \frac{1}{\sqrt{2}}\left(\a^{\dag\,i}_n\a^{\dag\,j}_{-n}+\a^{\dag\,j}_n\a^{\dag\,i}_{-n}
-\frac{1}{2}\d^{ij}\a^{\dag\,k}_n\a^{\dag\,k}_{-n}\right)|\a\ra \,,\label{9}
\end{align}
and analogously for $(i,j)\to(i',j')$. Here $\bf 3^{\pm}$ indicate the (anti)-selfdual representations. Under worldsheet-parity $\Omega$ (i.e. $n\leftrightarrow -n$) 
the singlets and the symmetric-traceless representations are even, whereas the (anti)-selfdual are odd. 
The ($\d$-function)-normalization\footnote{We will always suppress the $\d$-function normalization factor $|\a_3|\d(\a_3+\a_4)$, where $\a_3$ and $\a_4$ denote the
light-cone momenta of the in-/out-states, respectively.}
of $|\bf [9,1]\ra$ is $1+\frac{1}{2}\d^{ij}$, all other states are normalized to one. \\
Two fermionic oscillators lead to the states (cf. Appendix~\ref{appA} for our conventions for the $\s$-matrices)
\begin{align}
|{\bf [1,1]}\ra & = \frac{1}{2}\bigl(\b_n^{\dag}\bigr)_{\a_1\a_2}\bigl(\b_{-n}^{\dag}\bigr)^{\a_1\a_2}|\a\ra\,,\\
{|\bf[3^+,1]\ra}^{[ij]} & = \frac{1}{2}\bigl(\s^{ij}\bigr)^{\da_1\db_1}\bigl(\b_n^{\dag}\bigr)_{\da_1\da_2}\bigl(\b_{-n}^{\dag}\bigr)_{\db_1}^{\da_2}|\a\ra\,,\\
{|\bf[3^-,1]\ra}^{[ij]} & = \frac{1}{2}\bigl(\s^{ij}\bigr)^{\a_1\b_1}\bigl(\b_n^{\dag}\bigr)_{\a_1\a_2}\bigl(\b_{-n}^{\dag}\bigr)_{\b_1}^{\a_2}|\a\ra\,,\\
{|\bf [4,4]\ra}^{ij'}_{\pm} & = \frac{1\pm\Omega}{2\sqrt{2}}\bigl(\s^i\bigr)^{\da_1\a_1}\bigl(\s^{j'}\bigr)^{\da_2\a_2}
\bigl(\b_n^{\dag}\bigr)_{\a_1\a_2}\bigl(\b_{-n}^{\dag}\bigr)_{\da_1\da_2}|\a\ra\,,\\
{|\bf[3^-,3^-]\ra}^{[ij][i'j']} & = \frac{1}{2}\bigl(\s^{ij}\bigr)^{\a_1\b_1}\bigl(\s^{i'j'}\bigr)^{\a_2\b_2}
\bigl(\b_n^{\dag}\bigr)_{\a_1\a_2}\bigl(\b_{-n}^{\dag}\bigr)_{\b_1\b_2}|\a\ra\,,
\end{align}
and similarly for the remaining representations. Here, the singlets and $|\bf[3^{\pm},3^{\pm}]\ra$ have odd, whereas the (anti)-selfdual representations have even 
worldsheet-parity. Again we have normalized the states to one. 
Notice that only ${|\bf [9,1]\ra}^{(ij)}$, ${|\bf [1,9]\ra}^{(i'j')}$ and ${|\bf[3^{\pm},3^{\pm}]\ra}^{[ij][i'j']}$ are uniquely constructed of 
bosonic or fermionic oscillators, respectively. \\
The remaining representations are realized both with bosonic and fermionic oscillators and, therefore potentially mix with each 
other\footnote{In this case it is sometimes convenient to introduce $\bigl(\a_n^{\dag}\bigr)_{\a_1\da_1}\equiv\frac{1}{\sqrt{2}}\bigl(\s^i\bigr)_{\a_1\da_1}\a_n^{\dag\,i}$ and 
$\bigl(\a_n^{\dag}\bigr)_{\a_2\da_2}\equiv\frac{1}{\sqrt{2}}\bigl(\s^{i'}\bigr)_{\a_2\da_2}\a_n^{\dag\,i'}$. Then e.g.
${|\bf[3^+,1]\ra}^{[ij]} = \frac{1}{2}\bigl(\s^{ij}\bigr)^{\da_1\db_1}\bigl(\a_n^{\dag}\bigr)_{\a_1\da_1}\bigl(\a_{-n}^{\dag}\bigr)_{\db_1}^{\a_1}|\a\ra$.}.
For completeness we mention that fermionic two-impurity states transform as $\bf [(2,1),(1,2)]$, $\bf [(3,2),(2,1)]$, $\bf [(2,3),(1,2)]$ 
and the same representations with the two $SO(4)$'s exchanged. These are e.g.
\begin{align}
|{\bf [(2,1),(1,2)}\ra_{\pm\,\a_1\da_2} & = \frac{1\pm\Omega}{2}\bigl(\a_{-n}^{\dag}\bigr)_{\a_1\da_1}\bigl(\b_n^{\dag}\bigr)^{\da_1}_{\da_2}|\a\ra\,,\\
{|{\bf [(2,3),(1,2)}\ra}^{[ij]}_{\pm\,\a_1\da_2} & = \frac{1\pm\Omega}{2}\bigl(\s^{ij}\bigr)^{\da_1\db_1}
\bigl(\a_{-n}^{\dag}\bigr)_{\a_1\da_1}\bigl(\b_n^{\dag}\bigr)_{\db_1\da_2}|\a\ra\,,\\
{|{\bf [(3,2),(2,1)}\ra}^{[ij]}_{\pm\,\da_1\a_2} & = \frac{1\pm\Omega}{2}\bigl(\s^{ij}\bigr)^{\a_1\b_1}
\bigl(\a_{-n}^{\dag}\bigr)_{\a_1\da_1}\bigl(\b_n^{\dag}\bigr)_{\b_1\a_2}|\a\ra\,.
\end{align}
To construct the two-impurity multiplet explicitly we determine the highest weight state that is annihilated by $P^I$, $J^{I+}$, $Q^+$ and $\bar{Q}^+$ 
(no zero-mode excitations),  $J^{ij}$ and $J^{i'j'}$ (a singlet) and by half of the dynamical supercharges which we choose to be 
$\bar{Q}^-_{\a_1\da_2}$ and $Q^-_{\da_1\a_2}$ (cf. Appendix~\ref{appA} for their explicit oscillator expressions). The
latter requirements impose three conditions on the most general ansatz (with $|A|^2+|B|^2+|C|^2+|D|^2=1/4$ to normalize it to one)
\begin{equation}
|{\bf [1,1]}\ra_{(1)}  = 
\left(A\,\a_n^{\dag\,i}\a_{-n}^{\dag\,i}+B\,\a_n^{\dag\,i'}\a_{-n}^{\dag\,i'}+
C\,\bigl(\b_n^{\dag}\bigr)_{\a_1\a_2}\bigl(\b_{-n}^{\dag}\bigr)^{\a_1\a_2}+D\,\bigl(\b_n^{\dag}\bigr)_{\da_1\da_2}\bigl(\b_{-n}^{\dag}\bigr)^{\da_1\da_2}\right)|\a\ra\,,
\end{equation}
with the (up to a phase in $A$) unique solution
\begin{equation}
A = \frac{\omega_n-\m\a}{4\omega_n}\,,\qquad B = -\left(\frac{\omega_n+\m\a}{n}\right)^2A 
\,,\qquad C=-D=-ie(\a)\frac{\omega_n+\m\a}{n}A\,.
\end{equation}
As $|{\bf [1,1]}\ra_{(1)}$ contains states of opposite worldsheet parity, $\Omega$ is not a quantum number to label states in the supermultiplet;
in particular for large $\m$ and $\a<0$ we have 
\begin{equation}
A \sim \frac{1}{2}\,,\qquad B \sim -\frac{n^2}{8}\lambda'\,,\qquad C=-D \sim \frac{in}{4}\sqrt{\lambda'}\,,
\end{equation}
reflecting that in the BMN gauge theory the singlet operator built out of scalar impurities starts to mix with covariant derivatives and fermions at higher 
(than one) loops. 
By applying $Q_{\a_1\da_2}^-$ and $\bar{Q}_{\da_1\a_2}^-$ successively to $|{\bf [1,1]}\ra_{(1)}$ we generate all $2^8=256$ states with two stringy 
excitations. For example, acting twice with $\sqrt{\m}\,\d \equiv e^{\a_1\da_2}Q_{\a_1\da_2}^-+\bar{e}^{\da_1\a_2}\bar{Q}_{\da_1\a_2}^-$ we find
\footnote{We define
$e^{ij} = e_{\a_1\da_2}\bigl(\s^{ij}\bigr)^{\a_1\b_1}e^{\da_2}_{\b_1}\,,\quad
\bar{e}^{ij} = \bar{e}_{\da_1\a_2}\bigl(\s^{ij}\bigr)^{\da_1\db_1}\bar{e}^{\a_2}_{\db_1}\,,\quad
e^{ij'} = e_{\a_1\da_2}\bigl(\s^i\bigr)^{\da_1\a_1}\bigl(\s^{j'}\bigr)^{\da_2\a_2}\bar{e}_{\da_1\a_2}\,,$
and analogously for $e^{i'j'}$ etc.}
\begin{align}\label{susyvar}
\d^2|{\bf [1,1]}\ra_{(1)} & = \frac{ie(n)}{\m\a}\sqrt{8\omega_n(\omega_n+\m\a)}\Big(e^{ij}|{\bf [3^-,1]}\ra^{[ij]}_{(1)}-\bar{e}^{ij}|{\bf [3^+,1]}\ra^{[ij]}_{(1)}\Big)
-4\frac{\omega_n}{\m\a}e^{ij'}|{\bf [4,4]}\ra^{ij'}_{+(1)}\notag\\
& -\frac{ie(n)}{\m\a}\sqrt{8\omega_n(\omega_n-\m\a)}\Big(e^{i'j'}|{\bf [1,3^+]}\ra^{[i'j']}_{(2)}-\bar{e}^{i'j'}|{\bf [1,3^-]}\ra^{[i'j']}_{(2)}\Big)\,.
\end{align}
Here 
\begin{align}
|{\bf[3^+,1]}\ra^{[ij]}_{(1/2)} & = \bigl(\s^{ij}\bigr)^{\da_1\db_1}
\left(A_{(1/2)}\bigl(\a_n^{\dag}\bigr)_{\a_1\da_1}\bigl(\a_{-n}^{\dag}\bigr)_{\db_1}^{\a_1}
+B_{(1/2)}\bigl(\b_n^{\dag}\bigr)_{\da_1\da_2}\bigl(\b_{-n}^{\dag}\bigr)_{\db_1}^{\da_2}\right)|\a\ra\,,\\
|{\bf[3^-,1]}\ra^{[ij]}_{(1/2)} & = \bigl(\s^{ij}\bigr)^{\a_1\b_1}
\left(A_{(1/2)}\bigl(\a_n^{\dag}\bigr)_{\a_1\da_1}\bigl(\a_{-n}^{\dag}\bigr)_{\b_1}^{\da_1}
-B_{(1/2)}\bigl(\b_n^{\dag}\bigr)_{\da_1\da_2}\bigl(\b_{-n}^{\dag}\bigr)_{\b_1}^{\a_2}\right)|\a\ra\,,\\
|{\bf [4,4]}\ra^{ij'}_{+\,(1)} & = \frac{1+\Omega}{4}\bigl(\s^i\bigr)^{\da_1\a_1}\bigl(\s^{j'}\bigr)^{\da_2\a_2}
\left(\bigl(\a_n^{\dag}\bigr)_{\a_1\da_1}\bigl(\a_{-n}^{\dag}\bigr)_{\a_2\da_2}
+e(\a)\bigl(\b_n^{\dag}\bigr)_{\a_1\a_2}\bigl(\b_{-n}^{\dag}\bigr)_{\da_1\da_2}\right)|\a\ra\,,
\end{align}
and analogously for $(i,j)\to(i',j')$, while the mixing-coefficients are
\begin{equation}
A_{(1/2)} = \sqrt{\frac{\omega_n\mp\mu \a}{8 \omega_n}}\,,\qquad B_{(1/2)} = \pm \, i\, e(\a\, n) \sqrt{\frac{\omega_n\pm\mu \a}{8 \omega_n}}\,.
\end{equation} 
In the large $\m$ limit and $\a<0$ this yields e.g. $A_1\sim\frac{1}{2}$, $B_1\sim\frac{in}{2}\sqrt{\lambda'}$ so the mixing of states is again a $\sqrt{\lambda'}$ effect. 
For $|{\bf [4,4]}\ra_{+\,(1)}$ the mixing is maximal in agreement with the gauge theory result. Up to irrelevant phases and an overall factor of $\sqrt{2}$ 
(which can be absorbed in the definition of $e^{\a_1\da_2}$, $\bar{e}^{\da_1\a_2}$) the leading order SUSY variation \eqref{susyvar} precisely agrees 
with~\cite{Beisert:2002tn}.


\section{Turning on Interactions}\label{section3}

String interactions in the plane-wave background have been treated within the framework of 
light-cone string field theory~\cite{Spradlin:2002ar,Pankiewicz:2003kj,Chu:2002eu}. Its guiding principles are worldsheet continuity and the realization 
of the superalgebra in the full interacting theory: the superalgebra gives rise to two types of constraints -- 
kinematical and dynamical -- depending on whether the participating generators receive $g_2$ corrections ($H$, $Q^-$ and $\bar{Q}^-$) or not. 
Kinematical constraints lead to the continuity conditions in superspace, whereas dynamical constraints require the insertion of interaction point 
operators~\cite{Mandelstam:hk}. 
In practice these constraints will be solved perturbatively, for example $H$, the full Hamiltonian of the interacting theory, has an expansion 
in $g_2$ 
\begin{equation}\label{hexp}
H=H_2+g_2H_3+g_2^2H_4+\cdots\,.
\end{equation}
Here the operator $H_3$ represents a three-string interaction, but it is more convenient
to express it as a state $|H_3\ra$ in the multi-string Hilbert space and work in the number basis~\cite{Cremmer:1974jq}. 
Then the dynamical generators are of the form ${\mc P}|V\ra$, 
where ${\mc P}$ are the prefactors determined by the dynamical constraints (i.e. the oscillator expressions of the interaction point 
operators mentioned above) and the kinematical part of the vertex $|V\ra$ common to all 
the dynamical generators implements the continuity conditions.
These follow  for example from $[H_3,P^I]=0$, so the interaction vertex is translationally invariant and conserves transverse momentum. 
In the number basis the bosonic part of $|V\ra$ has the form 
\begin{equation}\label{bv}
|E_{\a}\ra = \exp\left(\frac{1}{2}
\sum_{s,t=1}^3\sum_{m,n\in\mathbb{Z}}\a^{\dag\,I}_{m(s)}\wt{N}^{st}_{mn}\a^{\dag\,I}_{n(t)}\right)|\a\ra_{123}\,,
\end{equation}
where $|\a\ra_{123}=|\a\ra_1\otimes|\a\ra_2\otimes|\a\ra_3$ is the tensor product of three (bosonic) vacuum states and $\wt{N}^{st}_{mn}$ 
are known as Neumann matrices, see~\cite{He:2002zu} for explicit expressions as functions of $\m$, $\a_s$. \\
To fulfil the dynamical constraints we define the linear combinations of the free supercharges $\sqrt{2}\eta\,Q\equiv Q^-+i\bar{Q}^-$ and 
$\sqrt{2}\bar{\eta}\,\wt{Q}=Q^--i\bar{Q}^-$ ($\eta=e^{i\pi/4}$) which satisfy on the space of physical states e.g. 
\begin{equation}
\begin{split}
\{Q_{\a_1\da_2},Q_{\b_1\db_2}\} & = \{\wt{Q}_{\a_1\da_2},\wt{Q}_{\b_1\db_2}\} = -2\e_{\a_1\b_1}\e_{\da_2\db_2}H\,,\\
\{Q_{\a_1\da_2},\wt{Q}_{\b_1\db_2}\} & = -\m\,\e_{\da_2\db_2}\bigl(\s^{ij}\bigr)_{\a_1\b_1}J^{ij}+\m\,\e_{\a_1\b_1}\bigl(\s^{i'j'}\bigr)_{\da_2\db_2}J^{i'j'}\,,
\end{split}
\end{equation}
and similar relations for $Q_{\da_1\a_2}$ and $\wt{Q}_{\db_1\b_2}$. Since $J^{ij}$ and $J^{i'j'}$ are not corrected by the interaction,
it follows that at order ${\mc O}(g_2)$ the dynamical generators have to obey
\begin{align}
\{Q_{2\,\a_1\da_2},Q_{3\,\b_1\db_2}\}+\{Q_{3\,\a_1\da_2},Q_{2\,\b_1\db_2}\} & = -2\e_{\a_1\b_1}\e_{\da_2\db_2}H_3\,,\label{dyn1}\\
\{\wt{Q}_{2\,\a_1\da_2},\wt{Q}_{3\,\b_1\db_2}\}+\{\wt{Q}_{3\,\a_1\da_2},\wt{Q}_{2\,\b_1\db_2}\} & = -2\e_{\a_1\b_1}\e_{\da_2\db_2}H_3\,,\label{dyn2}\\
\{Q_{2\,\a_1\da_2},\wt{Q}_{3\,\b_1\db_2}\} + \{Q_{3\,\a_1\da_2},\wt{Q}_{2\,\b_1\db_2}\} & = 0\label{dyn3}\,.
\end{align}
Substituting the most general ansatz for, say $Q_{3\,\a_1\da_2}$, compatible with  the requirement that the Hamiltonian prefactor in 
its functional form is quadratic in derivatives, into~\eqref{dyn1} and demanding that the result only involves the tensor $\e_{\a_1\b_1}\e_{\da_2\db_2}$
fixes $Q_{3\,\a_1\da_2}$ and consequently also $H_3$ up to their normalization. The same procedure applies to $\wt{Q}_{3\,\a_1\da_2}$ and requires that its normalization is the 
same as of $Q_{3\,\a_1\da_2}$.\\
In short, the three-string vertex and dynamical supercharges are 
\begin{align}
\label{H}
g_2|H_3\ra & = 
-g_2\,f(\m\a_3\,,\,\tfrac{\a_1}{\a_3})\frac{\k}{4\,\a_3^3}
\left[\sum_{s=1}^3\sum_{m\in\mathbb{Z}}\frac{\omega_{m(s)}}{\a_s}\a_{m(s)}^{\dag\,I}\a_{-m(s)}^J-\frac{\m}{2}\d^{IJ}\right]v_{IJ}|V\ra\,,\\
\label{Q1}
g_2|Q_{3\,\b_1\db_2}\ra & = -g_2\,\bar{\eta}\,f(\m\a_3\,,\,\tfrac{\a_1}{\a_3})\frac{1}{4\, \a_3^3}\,\sqrt{-\frac{\a'\k}{2}}\wt{K}^Iq_{I\b_1\db_2}|V\ra\,,\\
\label{Q2}
g_2|Q_{3\,\db_1\b_2}\ra & = -g_2\,\bar{\eta}\,f(\m\a_3\,,\,\tfrac{\a_1}{\a_3})\frac{1}{4\, \a_3^3}\,\sqrt{-\frac{\a'\k}{2}}\wt{K}^Iq_{I\db_1\b_2}|V\ra\,,
\end{align}
with similar expressions for $|\wt{Q}_3\ra$. Here $\k\equiv\a_1\a_2\a_3$, $\a_3<0$ for the incoming and $\a_{1,2}>0$ for the outgoing strings and 
$\wt{K}^I$ is defined in~\eqref{k}. Further we list 
the relevant parts of $v_{IJ}$, $q_{I\b_1\db_2}$ and $q_{I\db_1\b_2}$ in the next section, for complete expressions see e.g.~\cite{Pankiewicz:2003kj}.
In equations~\eqref{H}-\eqref{Q2} we suppressed the integrals over light-cone momenta
$\a_t$ and the $\d$-function normalization factor $|\a_3|\d\bigl(\sum_{t=1}^3\a_t\bigr)$.\\
More importantly, as alluded to above, the normalization of the dynamical generators is 
not fixed by the superalgebra at order ${\mc O}(g_2)$ and can be an arbitrary (dimensionless) function $f(\m\a_3\,,\,\tfrac{\a_1}{\a_3})$ of the light-cone 
momenta and $\m$ due to the fact that $P^+$ is a central element of the algebra. 
Indeed, it does not seem that further consistency conditions at higher orders in $g_2$ would allow to fix $f$. \\  
Now consider the constraints at order ${\mc O}(g_2^2)$. These are e.g.
\begin{align}
\{Q_{2\,\a_1\da_2},Q_{4\,\b_1\db_2}\}+\{Q_{4\,\a_1\da_2},Q_{2\,\b_1\db_2}\} + \{Q_{3\,\a_1\da_2},Q_{3\,\b_1\db_2}\} & = -2\e_{\a_1\b_1}\e_{\da_2\db_2}H_4\,,\label{dyn4}\\
\{\wt{Q}_{2\,\a_1\da_2},\wt{Q}_{4\,\b_1\db_2}\}+\{\wt{Q}_{4\,\a_1\da_2},\wt{Q}_{2\,\b_1\db_2}\} + \{\wt{Q}_{3\,\a_1\da_2},\wt{Q}_{3\,\b_1\db_2}\}& = 
-2\e_{\a_1\b_1}\e_{\da_2\db_2}H_4\,,\label{dyn5}\\
\{Q_{2\,\a_1\da_2},\wt{Q}_{4\,\b_1\db_2}\} + \{Q_{4\,\a_1\da_2},\wt{Q}_{2\,\b_1\db_2}\} + \{Q_{3\,\a_1\da_2},\wt{Q}_{3\,\b_1\db_2}\}& = 0\label{dyn6}\,
\end{align}
and have been analyzed in some detail in~\cite{Greensite:1986gv}. In particular, it was found that $\{Q_3,Q_3\}$ and $\{Q_3,\wt{Q}_3\}$ diverge in the
$(2 \to 2)\,$- strings channel leading in the latter case to the introduction of a non-vanishing $Q_4$ (and $\wt{Q}_4$). The new supercharges -- together with $\{Q_3,Q_3\}$ --
generate a contact term $H_4$ needed for finite scattering amplitudes. An analogue argumentation presumably holds for the $(1 \to 1)\,$- string channel. 
We have checked that for our calculations the above conditions are not violated if $Q_4$ is set to zero. Still, this constitutes only a necessary but clearly not sufficient 
requirement.
    

\section{Computing energy shifts in light-cone SFT}

To compute the energy shift of two-impurity states to leading order in $g_2$, the relevant part of the Hamiltonian is 
\begin{equation}
H = H_2 + g_2 H_3 + g_2^2 H_4\,,
\end{equation}
where the contact term $H_4$ acting in the single-string Hilbert space is induced by the cubic supercharges\footnote{At this point we neglect possible contributions of $Q_4$.} 
\begin{equation}
H_4 = \frac{1}{8}Q_{3\,\b_1\db_2}Q_3^{\b_1\db_2}+\frac{1}{8}Q_{3\,\db_1\b_2}Q_3^{\db_1\b_2}\,.
\end{equation}
As we have seen in section~\ref{section2}, there are generically several two-impurity eigenstates 
transforming in the same irreducible representation of $SO(4)\times SO(4)$; hence these will mix with each other and we have to use degenerate perturbation 
theory to compute their energy shift. The required formula for the energy shift is standard and reads 
\begin{equation}\label{shift}
\d E_n^{(2)}\la \varphi^A_n|0\ra = g_2^2\sum_B\Biggl[\sum_{C}\sum_{p \neq n}\frac{\la \varphi^A_n|H_3|\psi^C_p \ra \la \psi^C_p|H_3|\varphi^B_n\ra}{E_n^{(0)}-E_p^{(0)}}
+ \la \varphi^A_n|H_4|\varphi^B_n\ra\Biggr]
\la\varphi^B_n|0\ra\,,
\end{equation}
Here $A$ and $B$ label the degenerate one-string eigenstates, $\sum_{C}\sum_{p \neq n}|\psi^C_p \ra \la \psi^C_p|$ is the two-string projector and 
$|0\ra= \sum_A \la \varphi^A_n|0\ra\, |\varphi^A_n \ra$.
Thus, we essentially have to diagonalize the mixing matrix
\begin{equation}
M^{AB} = \la \varphi^A_n|H_3\bigl(E_n^{(0)}-H_2\bigr)^{-1}H_3+H_4|\varphi^B_n\ra\,.
\end{equation}
Since the theory is supersymmetric, this can be achieved by constructing supermultiplets:
Suppose we have constructed the complete supermultiplet by acting with eight supercharges, say 
$Q^-_{\a_1\da_2}$ and $\bar{Q}^-_{\da_1\a_2}$ on a highest weight state that is annihilated by the remaining charges. Now consider two states $|1\ra$ and $|2\ra$
carrying the same $SO(4)\times SO(4)$ quantum numbers, related by, say
\begin{equation}
|2\ra = Q^{-\,4}|1\ra\,,\qquad Q^{-\,4}\equiv Q^-_{\a_1\da_2}{Q^-}^{\da_2}_{\b_1}{Q^-}^{\a_1}_{\db_2}{Q^-}^{\b_1\db_2} 
\equiv -\bigl(Q^-\bigr)_{\a_1\da_2}\bigl(Q^{-\,3}\bigr)^{\a_1\da_2}\,.
\end{equation}
Therefore we have to show that the off-diagonal matrix elements of $M^{AB}$ vanish
\begin{equation}\label{id}
\la 1|M|2\ra = \la 1|M\,Q^{-\,4}|1\ra = \la 1|\bigl[\bigl(Q^-\bigr)_{\a_1\da_2},M\bigr]\bigl(Q^{-\,3}\bigr)^{\a_1\da_2}|1\ra \stackrel{!}{=}0\,,
\end{equation}
where we used that $\la 1|Q^-_{\a_1 \da_2}=0$. As a matter of fact, equation~\eqref{id} is a consequence of supersymmetry. Recall
$\bigl[Q^-,H\bigr] = 0$, so in particular to order $g_2^2$ we have the conditions
\begin{align}
\bigl[Q_2^-,H_2\bigr] & = 0\,,\\
\bigl[Q_2^-,H_3\bigr] & = -\bigl[Q_3^-,H_2\bigr]\,,\\
\bigl[Q_2^-,H_4\bigr] & = -\bigl[Q_3^-,H_3\bigr]-\bigl[Q_4^-,H_2\bigr]\,.\label{q2h4}
\end{align}
Applying this to equation~\eqref{id}, using $\bigl\{\bigl(Q_2^-\bigr)_{\a_1\da_2},\bigl(Q_3^-\bigr)^{\da_1\da_2}\bigr\}=0$ and 
\begin{equation}
\la 1|H_3\bigl(E_n^{(0)}-H_2\bigr)^{-1}\bigl[\bigl(Q_3^-\bigr)_{\a_1\da_2},H_2\bigr]\bigl(Q^{-\,3}_2\bigr)^{\a_1\da_2}|1\ra
=\la 1|H_3 \bigl(Q_3^-\bigr)_{\a_1\da_2}\bigl(Q^{-\,3}_2\bigr)^{\a_1\da_2}|1\ra\,,
\end{equation}
we find according to~\eqref{q2h4} and the Jacobi identity for $\{\bigl[Q_4^-,H_2\bigr],Q^{-\,3}_2\}$, that
\begin{equation}
\la 1|M|2\ra = \la 1|\Bigl(\bigl[\bigl(Q_3^-\bigr)_{\a_1\da_2},H_3\bigr]+\bigl[\bigl(Q_2^-\bigr)_{\a_1\da_2},H_4\bigr]\Bigr)\bigl(Q^{-\,3}_2\bigr)^{\a_1\da_2}|1\ra = 0\,.
\end{equation}
This proves the claim. We would like to stress that this proof is only valid when taking into account the full two-string projector, i.e. including 
impurity conserving {\em and} non-conserving channels. The same holds for the well-known statement that all states in a supermultiplet receive the same 
energy corrections. \\
In the following we will compute as an example the energy shift of the symmetric traceless state $|{\bf[9,1]}\ra^{(ij)}$. Due to the lack of adequate techniques we will 
restrict ourselves to the analysis of the impurity-conserving channel. Our result will further demonstrate the necessity of including the missing channels.   

\subsection{The energy shift of $|{\bf[9,1]}\ra^{(ij)}$}

As we saw in section~\ref{section2}, the state $|{\bf[9,1]}\ra^{(ij)}$ given in \eqref{9} has no counterpart built out of fermionic oscillators\footnote{The same holds for 
the $|{\bf[3^{\pm},3^{\pm}]}\ra^{(ij)}$ states on the ``fermionic'' side.} and thus, this case is comparatively easy to consider.  
The relevant expressions for the vertex are
\begin{equation}
v_{ij} = \d_{ij}\,, \quad q_{\b_1 \db_2}^l  =-iZ_{\dg_1\db_2}{(\s^l)}^{\dg_1}_{\b_1}\,, \quad q_{\db_1 \b_2}^l  = -Y_{\g_1\b_2}{(\s^l)}^{\g_1}_{\db_1}\,,
\end{equation}
while the impurity-conserving part of the two-string projector $\sum_{C}\sum_{p \neq n}|\psi^C_p \ra \la \psi^C_p|$ reads\footnote{Note, for the same reason as described 
above, fermionic oscillators do not contribute in this case to ${\bf 1}_{\text{2imp}}^{\rm loop}$ .}
\begin{equation} \label{projector}
\begin{split}
{\bf 1}_{\text{2imp}}^{\rm loop} =&
\int_0^1\frac{dr}{2r(1-r)}\Big[
\a_0^{\dag\,k}\,| \wt{\a}_1\ra\,\a_0^{\dag\,l}\,|\wt{\a}_2\ra\la \wt{\a}_2|\,\a_0^l\,\la \wt{\a}_1|\,\a_0^k 
+\sum_{p\in\mathbb{Z}}
\a_p^{\dag\,k}\,\a_{-p}^{\dag\,l}\,| \wt{\a}_1\ra\,| \wt{\a}_2\ra\la \wt{\a}_2|\,\la \wt{\a}_1|\,\a_{-p}^l\,\a_p^k\Big]\,,\\
{\bf 1}_{\text{2imp}}^{\rm cont} =&
\int_0^1\frac{dr}{r(1-r)}\Big[
\a_0^{\dag\,k}\,|\wt{\a}_1\ra\,\b_0^{\dag\,a}\,|\wt{\a}_2\ra\la \wt{\a}_2|\,\b_0^a\,\la \wt{\a}_1|\,\a_0^k
+\sum_{p\in\mathbb{Z}}
\a_p^{\dag\,k}\,\b_{-p}^{\dag\,a}\,|\wt{\a}_1\ra\,|\wt{\a}_2\ra\la \wt{\a}_2|\,\la \wt{\a}_1|\,\b_{-p}^a\,\a_p^k\Big]\,,
\end{split}
\end{equation}
where $Y$ and $Z$ denote the fermionic constituents of the prefactor (cf. \eqref{yz}) and we sum over $k$, $l$ and the $SO(8)$ spinor index $a$. 
Further we define $\wt{\a}_1 \equiv -\a_3r$ and $\wt{\a}_2 \equiv -\a_3 (1-r)$ using already the $\d$-function normalizations 
$\a_3^2\, \d(\wt{\a}_1 + \wt{\a}_2 + \a_3)\,\d(\wt{\a}_1 + \wt{\a}_2 + \a_3) = \a_3^2 \, \d(\wt{\a}_1 + \wt{\a}_2 + \a_3) \,\d(\a_3 + \a_4)$ from the two vertices; 
the factor $|\a_3| \d(\a_3 + \a_4)$ will be suppressed, as it is cancelled by the normalization of the external states. \\
In our notation, the oscillators act on the vacuum right next to them and therefore do {\em not} carry an extra index. This convention also circumvents potential 
double-countings, e.g. treating $\a_0^{\dag \, i}|p^+\ra \,|p^+\ra$ and $|p^+\ra \,\a_0^{\dag \, i}|p^+\ra$ as different states although they both correspond to the large
$J$ limit of the field theory operator ${\rm Tr}[\phi_i\, Z^{J/2}]\,{\rm Tr}[Z^{J/2}]$. \\
When calculating the matrix elements of $H_3$ and $Q_3$, one obtains two contributions for each matrix element according to the two possibilities of contracting 
the vacua of ${\bf 1}_{\text{2imp}}$ with $|\a_1\ra$ and $|\a_2\ra$ of the vertex, meaning that $\a_1 = -\a_3 r$ and $\a_2 = -\a_3 (1-r)$ in the first and vice 
versa in the second term. Both give the same result since the vertex is a symmetric function of the light-cone momenta. 
One finds  
\begin{equation}
\begin{split}
 ^{(ij)}\la{\bf[9,1]}|\,\la \wt{\a}_2|\,\a_0^l\,\la \wt{\a}_1|\,\a_0^k \,|H_3\ra = &
- 2\,r(1-r)  \Big( \frac{\omega_{n(3)}}{\a_3} + \mu \Big) \,\wt{N}^{31}_{n,0}\,\wt{N}^{32}_{n,0}\,\D^{ijkl} \\
 ^{(ij)}\la{\bf[9,1]}|\,\la \wt{\a}_2|\,\la \wt{\a}_1|\,\a_{-p}^l\,\a_p^k \,|H_3\ra = &
- 2\,r(1-r) \Big( \frac{\omega_{n(3)}}{\a_3} - \frac{\omega_{p(1)}}{\a_3 r}\Big) \,\wt{N}^{31}_{n,p}\,\wt{N}^{31}_{n,-p}\,\D^{ijkl} \\
\end{split}
\end{equation} 
for the matrix elements of $H_3$ and
\begin{equation}
\begin{split}
 ^{(ij)}\la{\bf[9,1]}|\,\la \wt{\a}_2|\,(\b_0)^{\ds_1 \ds_2}\,\la \wt{\a}_1|\,\a_0^k\,|Q_{3\,\b_1\db_2}\ra = &
- 2i\, \bar{C}\, G_{0(2)}\, \big(K_{n(3)} + K_{-n(3)}\big)\, \wt{N}^{31}_{n,0}\, \D^{ijkl}(\s^l)^{\ds_1}_{\b_1}\,\d^{\ds_2}_{\db_2}\\
 ^{(ij)}\la{\bf[9,1]}|\,\la \wt{\a}_2|\,\la \wt{\a}_1|\,(\b_{-p})^{\ds_1 \ds_2}\,\a_p^k\,|Q_{3\,\b_1\db_2}\ra = &
- 2i\, \bar{C}\, G_{|p|(1)}\, \big(K_{n(3)}\wt{N}^{31}_{n,p} + K_{-n(3)}\wt{N}^{31}_{n,-p}\big)\, \D^{ijkl}(\s^l)^{\ds_1}_{\b_1}\,\d^{\ds_2}_{\db_2}\\
 ^{(ij)}\la{\bf[9,1]}|\,\la \wt{\a}_2|\,(\b_0)^{\s_1 \s_2}\,\la \wt{\a}_1|\,\a_0^k\,|Q_{3\,\db_1\b_2}\ra = &
-  2\, \bar{C}\, G_{0(2)}\, \big(K_{n(3)} + K_{-n(3)}\big)\, \wt{N}^{31}_{n,0}\, \D^{ijkl}(\s^l)^{\s_1}_{\db_1}\,\d^{\s_2}_{\b_2}\\
 ^{(ij)}\la{\bf[9,1]}|\,\la \wt{\a}_2|\,\la \wt{\a}_1|\,(\b_{-p})^{\s_1 \s_2}\a_p^k\,|Q_{3\,\db_1\b_2}\ra = &
- 2\, \bar{C}\, G_{|p|(1)}\, \big(K_{n(3)}\wt{N}^{31}_{n,p} + K_{-n(3)}\wt{N}^{31}_{n,-p}\big)\, \D^{ijkl}(\s^l)^{\s_1}_{\db_1}\,\d^{\s_2}_{\b_2}
\end{split}
\end{equation}
for $Q_3$, where $\D^{ijkl} \equiv \frac{1}{\sqrt{2}}\big\{\d^{ik}\d^{jl}+\d^{il}\d^{jk}-\frac{1}{2}\d^{ij}\d^{kl}\big\}$ and 
$\bar{C} \equiv \frac{\bar{\eta}}{4} \, \sqrt{-\frac{\a'}{2\,\a_3^3}}\, \sqrt{r(1-r)}$~\footnote{The vectors $K_{n(t)}$ and $G_{n(t)}$ appear in the oscillator 
expansions of the bosonic/fermionic constituents $\wt{K}$, $Y$ and $Z$.}. Here we have fixed the normalization function $f=1$ such that
the gauge theory matrix element $\la m; 1-r|\la r|\wt{H}|n;1\ra$ and $_{123}\la \a| \a_{n}^1\, \a_{-n}^2\, \a_p^1\, \a_{-p}^2\,|H_3\ra$ agree at 
leading order.
It should be mentioned, that regardless of the form of $f$ the matrix elements necessarily fail to match at higher order~\cite{Spradlin:2003bw}. Thus, this choice seems to be
somewhat arbitrary. Moreover, the final result for the energy shift might be very sensitive to the $r$-dependence of $f$ due to the integral in the projector. \\ 
After plugging in all remaining expressions (cf. Appendix \ref{appA}) the complete result - up to exponential corrections - for the $|{\bf[9,1]}\ra^{(ij)}$ state 
reads\footnote{Note, that we have used $(\D^{ijkl})^2 = 1+ \frac{1}{2}\d^{ij}$ and $|\D^{ijkl}(\s^l)^{\s_1}_{\db_1}|^2 = 2 (1+\frac{1}{2}\d^{ij})$.}
\begin{equation} \label{sum}
\begin{split}
\frac{1}{\mu}\,\d E_n^{(2)}  
&\approx g_2^2 \int_0^1 dr\,\frac{\sin^4(\pi n r)}{8\, \pi^4 \,\omega_{n(3)}^2}\bigg\{\frac{n^2 + 4\, \mu \, \a_3 \,( \mu \, \a_3 - \omega_{n(3)})}{2\, r\, n^2\,\mu^2\, \a_3^2} \\
&\hspace*{3.8cm}
- \frac{r^2 (1-r)}{\mu\,\a_3}\sum_{p>0} \frac{n^2\, \omega_{p(1)} - 2 \,\mu^2\,\a_3^2\,(\omega_{p(1)}-r\,\omega_{n(3)})}{\omega_{p(1)}^2\,(\omega_{p(1)}-r\,\omega_{n(3)})^2} \bigg\} \\
&\approx g_2^2 \int_0^1 dr\,\frac{\sin^4(\pi n r)}{16\, \pi^4\, n^2 \,\omega_{n(3)}^3 \,\m^2\,\a_3^2}\Big\{ 
\big[n^2 (1 -2(1-r)\m\,\a_3) + 4\,\m\, \a_3 (\m\,\a_3-\omega_{n(3)})\big]\omega_{n(3)} \\
&\hspace*{1cm}-2\, n\, (1-r) \m\,\a_3 \big[ (2 \,\m^2\,\a_3^2 + \omega_{n(3)}^2)(\text{arccsch}(\tfrac{\m\,\a_3}{n})+\pi \cot(\pi n r)) 
+ n\,\pi^2\,r\, \omega_{n(3)}^2\text{csc}^2(\pi n r)\big]\Big\}
\end{split}
\end{equation}
where we have cancelled the normalization factor $1+ \frac{1}{2}\d^{ij}$ on both sides and the sum is computed in Appendix \ref{sums}\footnote{It should be 
mentioned, that it is not consistent to take $\mu$ to infinity {\em before} computing the sum as this leads to divergent series.}. Performing the integral one finds
\begin{equation}
\begin{split}\label{shifttrue}
\frac{1}{\mu}\,\d E_n^{(2)}  
&\approx \frac{g_2^2}{768\, \pi^4\, n^2\, \omega_{n(3)}^3\,\m^2\, \a_3^2}
\Big\{ n^2 \Big(18\, \omega_{n(3)} - \m\, \a_3 \big(9 + 2\, \omega_{n(3)} (9 + 4 \,\pi^2 \,\omega_{n(3)})\,\big)\Big)\\
&\hspace*{1cm}-3\,\m\, \a_3 \big(9 \,\m^2\,\a_3^2 - 24 \,\omega_{n(3)}\,\m\, \a_3 + 32 \,\omega_{n(3)}^2\big)
- 18\, n\,\m\, \a_3 (\omega_{n(3)}^2 + 2 \,\m^2\,\a_3^2)\, \text{arccsch}(\tfrac{\m \,\a_3}{n})\Big\}\\  
&= \frac{g_2^2}{4\, \pi^2}\Big\{ \Big(\frac{1}{24}+\frac{65}{64\, n^2\pi^2}\Big) \lambda' -\frac{3}{16\, \pi^2} {\lambda'}^{3/2} 
- \frac{\,n^2}{2}\Big(\frac{1}{24}+\frac{89}{64\, n^2\pi^2}\Big)   {\lambda'}^{2} + \frac{9\, n^2}{32\, \pi^2}{\lambda'}^{5/2}+\cdots \Big\}\,.
\end{split}
\end{equation}
This result leads to several interesting observations, which we will discuss in what follows:
First of all, the leading order contribution in $\lambda'$ does {\em not} match the anomalous dimension computed in field theory (see~\cite{Beisert:2002bb,Beisert:2003tq}) and 
also disagrees with the original computation of~\cite{Roiban:2002xr}. The latter mismatch can be traced back to a reflection-symmetry factor $\frac{1}{2}$ introduced in 
equation (3.3) of~\cite{Roiban:2002xr}. Since the formula for the energy shift in equation~\eqref{shift} is standard quantum mechanical perturbation theory, such a factor
cannot be justified. Further we have been careful to define the two-string projectors -- concerning accidental over-counting -- and therefore we are confident that our result is
correct. \\
Even more, when repeating the analogous calculation for the states $|[{\bf 3^{\pm},1}]\ra^{[ij]}$, one does not obtain the result given in~\eqref{shifttrue} already at leading 
order in $\lambda'$. 
This computation coincides -- up to a factor of $\frac{1}{2}$ -- with that of the representation {\bf 6} done in~\cite{Roiban:2002xr}\footnote{Here only the contact term 
contributes and thus, the reflection-symmetry factor has no effect. The mismatch by $\frac{1}{2}$ is due to the different definition of the projectors 
(no sums over $r=1$, $2$), cf. the discussion below~\eqref{projector}.}, because the mixing with states consisting of fermionic oscillators starts to effect the energy shift 
at order ${\lambda'}^2$. 
Since both representations are in the same multiplet (cf. section \ref{section2}) and therefore are guaranteed to receive the same energy shift,   
this fact clearly shows, that it is not sufficient to restrict oneself only to the impurity-conserving channel.  
Moreover, additional contributions of a $Q_4$-induced contact term cannot be ruled out so far.
For ${\lambda'}^2$ the discrepancies increase. Now, not only the coefficients, but also the $n$-dependence do not reproduce the field theory result~\cite{Beisert:2003tq}.\\
A new interesting aspect is the appearance of half-integer powers of $\lambda'$. These obviously do not have a counterpart in perturbative, non-planar SYM.
The existence of half-integer powers in the expansion of Neumann matrix elements have already been noticed in~\cite{Klebanov:2002mp}, but to our knowledge this 
constitutes the first example of their appearance in a physical quantity. Although these might be an artefact of the truncation to the impurity-conserving channel, 
we expect them 
to be a generic feature of the complete SFT result representing a qualitative difference to the planar sector. \\  
In general one can say, that light-cone SFT is not sufficiently developed yet to give an unambiguous test of the non-planar part of the BMN correspondence. 
Apart from the presence of $Q_4$, which remains to be clarified, this mainly concerns the issue of normalization functions of the vertices not being fixed by the
superalgebra. It would desirable to find a SYM independent method to determine these.

\vskip1cm
\section*{Acknowledgements}
\vskip0.2cm
We are grateful to G.~Arutyunov, N.~Beisert, J.~Plefka, M.~Staudacher and B.~Stefa\'{n}ski jr. for useful discussions.
AP acknowledges support by the DOE under contract DE-FGO2-96ER40956. AP and PG thank the 
Albert-Einstein Institute and the University of Washington respectively, for hospitality 
during the course of this project. 


\appendix

\section{Notations and Definitions}\label{appA}

In this Appendix we collect our conventions and definitions as well as some useful identities. 
Compared to~\cite{Spradlin:2002ar} we performed the following redefinitions of our oscillator basis for $n>0$ to have the standard level-matching condition
\begin{align}
\sqrt{2}a_n^i & \equiv \a_n^i+\a_{-n}^i\,, & i\sqrt{2}a_{-n}^i &\equiv \a_n^i-\a_{-n}^i\,, & a_0^i &\equiv \a_0^i\,,\\
\sqrt{2}a_n^{i'} & \equiv \a_n^{i'}+\a_{-n}^{i'}\,, &  i\sqrt{2}a_{-n}^{i'} &\equiv \a_n^{i'}-\a_{-n}^{i'}\,, & a_0^{i'} &\equiv \a_0^{i'}\,,\\
\sqrt{2}b_n^{\a_1\a_2} & \equiv \b_n^{\a_1\a_2}+\b_{-n}^{\a_1\a_2}\,,& 
i\sqrt{2}b_{-n}^{\a_1\a_2} &\equiv \b_n^{\a_1\a_2}-\b_{-n}^{\a_1\a_2}\,,&
b_0^{\a_1\a_2} &\equiv \b_0^{\a_1\a_2}\,,\\
i\sqrt{2}b_n^{\da_1\da_2} & \equiv -\b_n^{\da_1\da_2}+\b_{-n}^{\da_1\da_2}\,,&
\sqrt{2}b_{-n}^{\da_1\da_2} &\equiv \b_n^{\da_1\da_2}+\b_{-n}^{\da_1\da_2}\,,&
b_0^{\da_1\da_2} &\equiv \b_0^{\da_1\da_2}\,.
\end{align}
The oscillators obey the standard commutation relations\footnote{Note that e.g. $\bigl[{\b_n}_{\a_1}^{\a_2}\bigr]^{\dag}=-{\b_n^{\dag}}^{\a_1}_{\a_2}$.}  
\begin{equation}
[\a_m^i,\a_n^{\dag\,j}]  = \d_{mn}\d^{ij}\,, \quad \{\bigl(\b_m\bigr)_{\a_1\a_2},\bigl(\b_n^{\dag}\bigr)^{\b_1\b_2}\} = \d_{mn}\d^{\b_1}_{\a_1}\d^{\b_2}_{\a_2}\,.
\end{equation}
It is convenient to introduce
\begin{equation}
\bigl(\a_n^{\dag}\bigr)_{\a_1\da_1}  \equiv \frac{1}{\sqrt{2}}\bigl(\s^i\bigr)_{\a_1\da_1}\a_n^{\dag\,i}\,, \quad
\bigl(\a_n^{\dag}\bigr)_{\a_2\da_2}  \equiv \frac{1}{\sqrt{2}}\bigl(\s^{i'}\bigr)_{\a_2\da_2}\a_n^{\dag\,i'} 
\end{equation}
which satisfy 
\begin{equation}
[\bigl(\a_m\bigr)_{\a_1\da_1},\bigl(\a_n^{\dag}\bigr)^{\db_1\b_1}] = \d_{mn}\d^{\b_1}_{\a_1}\d^{\db_1}_{\da_1}\,,\qquad
[\bigl(\a_m\bigr)_{\a_2\da_2},\bigl(\a_n^{\dag}\bigr)^{\db_2\b_2}] = \d_{mn}\d^{\b_2}_{\a_2}\d^{\db_2}_{\da_2}\,.
\end{equation}
The $\s$ matrices consist of the usual Pauli-matrices together with the $2d$ unit matrix 
(spinorial indices are raised and lowered with $\e_{\a\b}=\e_{\da\db}\equiv\bigl(\begin{smallmatrix} 0 & 1 \\ -1 & 0 \end{smallmatrix} \bigr)$) 
$\s^i_{\a\da} =\bigl(i\t^1,i\t^2,i\t^3,-1\bigr)_{\a\da}$\,. 
Notice the reality properties $\bigl[\s^i_{\a\da}\bigr]^{\dag} = {\s^i}^{\da\a}$\,, 
$\bigl[{\s^i}_{\a}^{\da}\bigr]^{\dag} = -{\s^i}^{\a}_{\da}$
which are consistent with the above commutation relations. Some useful identities are 
\begin{align}
\e_{\a\b}\e^{\g\d} & = \d_{\a}^{\d}\d_{\b}^{\g}-\d_{\a}^{\g}\d_{\b}^{\d}\,,\\
\s^i_{\a\db}{\s^j}^{\db}_{\b} & = -\d^{ij}\e_{\a\b}+\s^{ij}_{\a\b}\,,\qquad
(\s^{ij}_{\a\b}\equiv \s^{[i}_{\a\da}{\s^{j]}}^{\da}_{\b}=\s^{ij}_{\b\a})\\
\s^i_{\a\da}{\s^j}^{\a}_{\db} & = -\d^{ij}\e_{\da\db}+\s^{ij}_{\da\db}\,,
\qquad (\s^{ij}_{\da\db}\equiv \s^{[i}_{\a\da}{\s^{j]}}^{\a}_{\db}=\s^{ij}_{\db\da})\\
\s^k_{\a\da}\s^{k}_{\b\db} & = 2\e_{\a\b}\e_{\da\db}\,,\\
\frac{1}{2}\e^{ijkl}\s^{kl}_{\a\b} & = -\s^{ij}_{\a\b}\,,\\
\frac{1}{2}\e^{ijkl}\s^{kl}_{\da\db} & = \s^{ij}_{\da\db}\,.
\end{align}
The free dynamical supercharges are 
\begin{align}
\sqrt{\frac{|\a|}{2}}Q^-_{\a_1\da_2} & = -\frac{\sqrt{\m|\a|}}{2\sqrt{2}}(1-e(\a))
\Bigl[\a_{0\,\a_1}^{\,\,\,\,\db_1}\b_{0\,\db_1\da_2}^{\dag}+\a_{0\,\da_2}^{\dag\,\b_2}\b_{0\,\a_1\b_2}\Bigr]\notag\\
&+\sum_{k\neq 0}
\Biggl[\sqrt{\omega_k+\m\a}\,\a^{\dag\,\db_1}_{k\,\a_1}\b_{k\,\db_1\da_2}-ie(\a k)\sqrt{\omega_k-\m\a}\,\a_{k\,\a_1}^{\,\,\,\,\db_1}\b_{k\,\db_1\da_2}^{\dag}\notag\\
&-e(\a)\Bigl(\sqrt{\omega_k+\m\a}\,\a_{k\,\da_2}^{\,\,\,\,\b_2}\b_{k\,\a_1\b_2}^{\dag}-ie(\a k)\sqrt{\omega_k-\m\a}\,\a_{k\,\da_2}^{\dag\,\b_2}\b_{k\,\a_1\b_2}^{\dag}\Bigr)\Biggr]\,,\\
\sqrt{\frac{|\a|}{2}}Q^-_{\da_1\a_2} & = \frac{\sqrt{\m|\a|}}{2\sqrt{2}}(1+e(\a))
\Bigl[\a_{0\,\da_1}^{\,\,\,\,\b_1}\b_{0\,\b_1\a_2}^{\dag}+\a_{0\,\a_2}^{\dag\,\db_2}\b_{0\,\da_1\db_2}\Bigr]\notag\\
&+\sum_{k\neq 0}
\Biggl[\sqrt{\omega_k+\m\a}\,\a^{\dag\,\db_2}_{k\,\a_2}\b_{k\,\da_1\db_2}-ie(\a k)\sqrt{\omega_k-\m\a}\,\a_{k\,\a_2}^{\,\,\,\,\db_2}\b_{k\,\da_1\db_2}^{\dag}\notag\\
&+e(\a)\Bigl(\sqrt{\omega_k+\m\a}\,\a_{k\,\da_1}^{\,\,\,\,\b_1}\b_{k\,\b_1\a_2}^{\dag}-ie(\a k)\sqrt{\omega_k-\m\a}\,\a_{k\,\da_1}^{\dag\,\b_1}\b_{k\,\b_1\a_2}^{\dag}\Bigr)\Biggr]\,,
\end{align}
and $\bar{Q}^-=e(\a)\bigl[Q^-\bigr]^{\dag}$. \\
Now we present the quantities appearing in the vertex; namely Neumann matrices, bosonic and fermionic prefactors and all related functions.
Note, that for simplicity will already set $\b_1\equiv r$ and $\b_2 \equiv 1-r$ (with $\b_t\equiv-\a_t/\a_3$ and $\a_3 < 0$) in some of the expressions. 
The Neumann matrices appearing in the bosonic vertex read
\begin{equation}
\wt{N}^{st}_{mn}=
\begin{cases}
\frac{1}{2}\bar{N}^{st}_{|m||n|}\bigl(1+U_{m(s)}U_{n(t)}\bigr) & \,,m,n\neq0 \\
\frac{1}{\sqrt{2}}\bar{N}^{st}_{|m|0} & \,, m\neq0 \\
\bar{N}^{st}_{00}\,,
\end{cases}
\end{equation}
with\footnote{To have a manifest symmetry in $1\leftrightarrow 2$ we additionally redefined the oscillators as $(-1)^{s(n+1)}\a_{n(s)} \to \a_{n(s)}$ for 
$n\in\mathbb{Z}$, $s=1,2,3$ and analogously for the fermionic oscillators.}
\begin{align}
\label{mn}
\bar{N}^{st}_{mn} & =-(1-4\m\k K)^{-1}\frac{\k}{\a_s\omega_{n(t)}+\a_t\omega_{m(s)}}\left[CU_{(s)}^{-1}C_{(s)}^{1/2}\bar{N}^s\right]_m
\left[CU_{(t)}^{-1}C_{(t)}^{1/2}\bar{N}^t\right]_n\,,\\
\label{m0} \bar{N}^{st}_{m0} & =
\sqrt{-2\m\k(1-\b_t)}\sqrt{\omega_{m(s)}}\bar{N}^s_m\,,\qquad t\in\{1,2\}\,,\\
\label{00a} \bar{N}^{st}_{00} & =
(1-4\m\k K)\left(\d^{st}-\sqrt{\b_s\b_t}\right)\,,\qquad  s,t\in\{1,2\}\,,\\
\label{00b} \bar{N}^{s3}_{00} & = -\sqrt{\b_s}\,,\qquad s\in\{1,2\}\,.
\end{align}
Here we have used the short cuts 
\begin{align}
C_n & = n\,,\qquad C_{n(s)} = \omega_{n(s)}\equiv\sqrt{n^2+\bigl(\m\a_s\bigr)^2}\,,\qquad \k\equiv \a_1\a_2\a_3\,,\\
U_{n(s)} & =\frac{1}{n}(\omega_{n(s)}-\m\a_s)\,,\qquad U^{-1}_{n(s)}=\frac{1}{n}(\omega_{n(s)}+\m\a_s)\,.
\end{align}
Neglecting exponential corrections $\sim{\mc O}(e^{-\m\a_3})$ the exact $\m$ dependence of the Neumann vectors $\bar{N}^s$ and scalar $K$ 
is~\cite{He:2002zu}\footnote{To compare with the definition used in~\cite{He:2002zu} note that 
$\bar{N}^s_{n\,\text{here}}=(-1)^{s(n+1)}U_{n(s)}C_{n(s)}^{-1/2}\bar{N}^s_{n\,\text{there}}$.}
\begin{align}
\label{K}
1-4\m\k K & \approx -\frac{1}{4\pi r(1-r)\m\a_3}\,,\\
\label{N3}
\a_3\bar{N}^3_n & \approx -\frac{\sin(n\pi r)}{\pi r(1-r)}\frac{1}{\omega_{n(3)}\sqrt{-2\m\a_3(\omega_{n(3)}+\m\a_3)}}\,,\\
\label{Nr}
\a_3\bar{N}^s_n & \equiv \a_3\bar{N}_n(\b_s) \approx -\frac{\sqrt{\b_s}}{2\pi r(1-r)}\frac{1}{\omega_{n(s)}\sqrt{-2\m\a_3(\omega_{n(s)}-\m\a_3\b_s)}}\,.
\end{align}
The bosonic constituents of the prefactor are defined as
\begin{align} \label{k}
K^I & = \sum_{s=1}^3\sum_{n\in\mathbb{Z}}K_{n(s)}\a_{n(s)}^{I\,\dagger}\,, & 
\wt{K}^I & = \sum_{s=1}^3\sum_{n\in\mathbb{Z}}K_{n(s)}\a_{-n(s)}^{I\,\dagger}\,,
\end{align}
with ($n\neq 0$)
\begin{align}
K_{0(s)} & = (1-4\m\k K)^{1/2}\sqrt{-\frac{2\m\k}{\a'}\bigl(1-\b_s\bigr)}\,,\qquad K_{0(3)} = 0\,,\\
K_{n(s)} & = -\frac{\k}{\sqrt{2\a'}\a_s}(1-4\m\k K)^{-1/2}(\omega_{n(s)}+\m\a_s)\sqrt{\omega_{n(s)}}\bar{N}^s_{|n|}\bigl(1-U_{n(s)}\bigr)\,,
\end{align}
while the fermionic constituents of the prefactor are 
\begin{equation} \label{yz}
Y^{\a_1\a_2} = \sum_{s=1}^3\sum_{n\in\mathbb{Z}}G_{|n|(s)}\b^{\dag\,\a_1\a_2}_{n(s)}\,,\qquad
Z^{\da_1\da_2} = \sum_{s=1}^3\sum_{n\in\mathbb{Z}}G_{|n|(s)}\b^{\dag\,\da_1\da_2}_{n(s)}\,,
\end{equation}
with ($n\neq 0$)
\begin{align}
G_{0(s)} & = (1-4\m\k K)^{1/2}\sqrt{1-\b_s}\,,\qquad G_{0(3)} = 0\,,\\
G_{n(s)} & = \frac{e(\a_s)}{\sqrt{2|\a_s|}}\frac{\sqrt{-\k}}{(1-4\m\k K)^{1/2}}\sqrt{(\omega_{n(s)}+\m\a_s)\omega_{n(s)}}\bar{N}^s_{|n|}\,.
\end{align}

\section{Sums}\label{sums}

The sum over $p$ in equation~\eqref{sum} splits up into two parts, namely
\begin{equation}
\sum_{p>0} \Big\{ n^2 \frac{1}{\omega_{p(1)}\,(\omega_{p(1)}-r\,\omega_{n(3)})^2} - 2\, \mu^2\,\a_3^2\,\frac{1}{\omega_{p(1)}^2\,(\omega_{p(1)}-r\,\omega_{n(3)})}\Big\}
:=  n^2\, S(\mu) - 2\,\mu^2\, \a_3^2\, T(\mu).
\end{equation}
In the remainder of this section we will use $a \equiv -\m \a_3 r$ and $b \equiv r\, \omega_{n(3)}$. Note, that $b>a\ge0$.  \\
For the computation of $S(\m)$ we use the following trick
\begin{align}
\sum_{p>0}\frac{1}{\sqrt{p^2+a^2}\bigl(\sqrt{p^2+a^2}-b\bigr)^2}
&=\frac{d}{db}\sum_{p>0}\frac{\sqrt{p^2+a^2}+b}{\sqrt{p^2+a^2}(p^2-b^2+a^2)} \notag\\
&=\frac{d}{db}\sum_{p>0}\left[\frac{1}{p^2-b^2+a^2}+\frac{b}{\sqrt{p^2+a^2}\bigl(p^2-b^2+a^2\bigr)}\right]\,,
\end{align}
where we have integrated over b and expanded with $\sqrt{p^2+a^2}+b$. The first part gives
\begin{equation}
\sum_{p>0}\frac{1}{p^2-b^2+a^2} = \frac{1-\sqrt{b^2-a^2}\pi\cot\bigl(\sqrt{b^2-a^2}\pi\bigr)}{2\bigl(b^2-a^2\bigr)}\,,
\end{equation}
while the second can be represented as (using the contour integral method, see e.g.~\cite{He:2002zu})
\begin{equation}\label{S}
\sum_{p>0}\frac{b}{\sqrt{p^2+a^2}\bigl(p^2-b^2+a^2\bigr)} = \frac{b}{2a(b^2-a^2)}-\frac{\pi}{2}\frac{\cot\bigl(\sqrt{b^2-a^2}\pi\bigr)}{\sqrt{b^2-a^2}}
+b\int_1^{\infty}\frac{dx\,\coth\bigl(a\pi x\bigr)}{\bigl(a^2x^2+b^2-a^2\bigr)\sqrt{x^2-1}}\,.
\end{equation}
We rewrite the integral as
\begin{align}
&\int_1^{\infty}\frac{dx\,\coth\bigl(a\pi x\bigr)}{\bigl(a^2x^2+b^2-a^2\bigr)\sqrt{x^2-1}} \notag\\
&=\int_1^{\infty}dx \bigg[\frac{1}{\bigl(a^2x^2+b^2-a^2\bigr)\sqrt{x^2-1}}-\frac{2}{a^2\bigl(x^2+\frac{b^2}{a^2}-1\bigr)\sqrt{x^2-1}\bigl(1-e^{2a\pi x}\bigr)}\bigg]
\approx \frac{\text{arccsch}\Bigl[\frac{a}{\sqrt{b^2-a^2}}\Bigr]}{b\sqrt{b^2-a^2}}\,,
\end{align}
where we utilized that for large $a$, $\frac{b^2}{a^2}$ finite, the second integral can be omitted, thereby again neglecting exponential corrections. 
\text{arccsch}$(x)$ is the inverse hyperbolic cosecans function. Hence we find
\begin{align}
\sum_{p>0}\frac{1}{\sqrt{p^2+a^2}\bigl(\sqrt{p^2+a^2}-b\bigr)^2}\, \approx&\,
\frac{d}{db}\Biggl[\frac{1}{2a(b-a)}-\pi\frac{\cot\bigl(\sqrt{b^2-a^2}\pi\bigr)}{\sqrt{b^2-a^2}}+\frac{\text{arccsch}\Bigl[\frac{a}{\sqrt{b^2-a^2}}\Bigr]}{b\sqrt{b^2-a^2}}\Biggr]
\notag\\
=&-\frac{1}{2a(b-a)^2}+\frac{1}{b^2-a^2}+\frac{b}{\bigl(b^2-a^2\bigr)^{3/2}}\Biggl[-\text{arccsch}\Bigl[\frac{a}{\sqrt{b^2-a^2}}\Bigr]\notag\\
&+\pi\cot\bigl(\sqrt{b^2-a^2}\pi\bigr) +\pi^2\sqrt{b^2-a^2}\csc^2\bigl(\sqrt{b^2-a^2}\pi\bigr)\Biggr]\,.
\end{align}
\noindent
Now we turn to the sum $T(\mu)$. It takes the form
\begin{equation}
\sum_{p>0}\frac{1}{\bigl(p^2+a^2\bigr)\bigl(\sqrt{p^2+a^2}-b\bigr)} = \sum_{p>0}\Biggl[\frac{1}{\sqrt{p^2+a^2}\bigl(p^2-b^2+a^2\bigr)}
+\frac{b}{\bigl(p^2+a^2\bigr)\bigl(p^2-b^2+a^2\bigr)}\Biggr]\,,
\end{equation}
where the first part was already computed in equation~\eqref{S}. The remaining sum yields
\begin{equation}
\sum_{p>0}\frac{1}{\sqrt{p^2+a^2}\bigl(p^2-b^2+a^2\bigr)} = \frac{1}{2a^2\bigl(b^2-a^2\bigr)}-\frac{\pi}{2b^2}\frac{\cot\bigl(\sqrt{b^2-a^2}\pi\bigr)}{\sqrt{b^2-a^2}}
-\frac{\pi}{2ab^2}\coth\bigl(\pi a\bigr)\,,
\end{equation}
and therefore up to exponential corrections one finds
\begin{equation}
\sum_{p>0}\frac{1}{\bigl(p^2+a^2\bigr)\bigl(\sqrt{p^2+a^2}-b\bigr)} \approx 
\frac{1}{2a^2(b-a)}-\frac{\pi}{b}\frac{\cot\bigl(\sqrt{b^2-a^2}\pi\bigr)}{\sqrt{b^2-a^2}}+\frac{\text{arccsch}\Bigl[\frac{a}{\sqrt{b^2-a^2}}\Bigr]}{b\sqrt{b^2-a^2}}\,.
\end{equation}



\begin{thebibliography}{99}

\bibitem{Berenstein:2002jq}
D.~Berenstein, J.~M.~Maldacena and H.~Nastase,
``Strings in flat space and pp waves from N = 4 super Yang Mills,''
JHEP {\bf 0204} (2002) 013
[arXiv:hep-th/0202021].

\bibitem{Blau:2001ne}
M.~Blau, J.~Figueroa-O'Farrill, C.~Hull and G.~Papadopoulos,
``A new maximally supersymmetric background of IIB superstring theory,''
JHEP {\bf 0201}, 047 (2002)
[arXiv:hep-th/0110242].

\bibitem{Metsaev:2001bj}
R.~R.~Metsaev,
``Type IIB Green-Schwarz superstring in plane wave Ramond-Ramond  background,''
Nucl.\ Phys.\ B {\bf 625}, 70 (2002)
[arXiv:hep-th/0112044].

\bibitem{Spradlin:2002ar}
M.~Spradlin and A.~Volovich,
``Superstring interactions in a pp-wave background,''
Phys.\ Rev.\ D {\bf 66} (2002) 086004
[arXiv:hep-th/0204146].
%
M.~Spradlin and A.~Volovich,
``Superstring interactions in a pp-wave background. II,''
JHEP {\bf 0301} (2003) 036
[arXiv:hep-th/0206073].
%
A.~Pankiewicz,
``More comments on superstring interactions in the pp-wave background,''
JHEP {\bf 0209} (2002) 056
[arXiv:hep-th/0208209].
%
A.~Pankiewicz and B.~Stefa\'{n}ski~jr.,
``pp-wave light-cone superstring field theory,''
Nucl.\ Phys.\ B {\bf 657} (2003) 79
[arXiv:hep-th/0210246].

\bibitem{Pankiewicz:2003kj}
A.~Pankiewicz,
``An alternative formulation of light-cone string field theory on the  plane wave,''
JHEP {\bf 0306} (2003) 047
[arXiv:hep-th/0304232].
%
A.~Pankiewicz and B.~Stefa\'{n}ski~jr.,
``On the uniqueness of plane-wave string field theory,''
arXiv:hep-th/0308062.
%
\bibitem{Chu:2002eu}
C.~S.~Chu, V.~V.~Khoze, M.~Petrini, R.~Russo and A.~Tanzini,
``A note on string interaction on the pp-wave background,''
Class.\ Quant.\ Grav.\  {\bf 21} (2004) 1999
[arXiv:hep-th/0208148].
%
C.~S.~Chu, M.~Petrini, R.~Russo and A.~Tanzini,
``String interactions and discrete symmetries of the pp-wave background,''
Class.\ Quant.\ Grav.\  {\bf 20} (2003) S457
[arXiv:hep-th/0211188].
%
P.~Di Vecchia, J.~L.~Petersen, M.~Petrini, R.~Russo and A.~Tanzini,
``The 3-string vertex and the AdS/CFT duality in the pp-wave limit,''
Class.\ Quant.\ Grav.\  {\bf 21} (2004) 2221
[arXiv:hep-th/0304025].
%
S.~Dobashi and T.~Yoneya,
``Resolving the Holography in the Plane-Wave Limit of AdS/CFT Correspondence,''
arXiv:hep-th/0406225.

\bibitem{Constable:2002hw}
N.~R.~Constable, D.~Z.~Freedman, M.~Headrick, S.~Minwalla, L.~Motl, A.~Postnikov and W.~Skiba,
``PP-wave string interactions from perturbative Yang-Mills theory,''
JHEP {\bf 0207} (2002) 017
[arXiv:hep-th/0205089].
%
C.~Kristjansen, J.~Plefka, G.~W.~Semenoff and M.~Staudacher,
``A new double-scaling limit of N = 4 super Yang-Mills theory and PP-wave strings,''
Nucl.\ Phys.\ B {\bf 643} (2002) 3
[arXiv:hep-th/0205033].

\bibitem{Gross:2002su}
D.~J.~Gross, A.~Mikhailov and R.~Roiban,
``Operators with large R charge in N = 4 Yang-Mills theory,''
Annals Phys.\  {\bf 301} (2002) 31
[arXiv:hep-th/0205066].

\bibitem{Beisert:2003tq}
N.~Beisert, C.~Kristjansen and M.~Staudacher,
``The dilatation operator of N = 4 super Yang-Mills theory,''
Nucl.\ Phys.\ B {\bf 664} (2003) 131
[arXiv:hep-th/0303060].

\bibitem{Santambrogio:2002sb}
A.~Santambrogio and D.~Zanon,
``Exact anomalous dimensions of N = 4 Yang-Mills operators with large R charge,''
Phys.\ Lett.\ B {\bf 545}, 425 (2002)
[arXiv:hep-th/0206079].

\bibitem{Beisert:2004hm}
N.~Beisert, V.~Dippel and M.~Staudacher,
``A novel long range spin chain and planar N = 4 super Yang-Mills,''
arXiv:hep-th/0405001.

\bibitem{Beisert:2002bb}
N.~Beisert, C.~Kristjansen, J.~Plefka, G.~W.~Semenoff and M.~Staudacher,
``BMN correlators and operator mixing in N = 4 super Yang-Mills theory,''
Nucl.\ Phys.\ B {\bf 650} (2003) 125
[arXiv:hep-th/0208178].
%
N.~R.~Constable, D.~Z.~Freedman, M.~Headrick and S.~Minwalla,
``Operator mixing and the BMN correspondence,''
JHEP {\bf 0210} (2002) 068
[arXiv:hep-th/0209002].

\bibitem{Beisert:2002ff}
N.~Beisert, C.~Kristjansen, J.~Plefka and M.~Staudacher,
``BMN gauge theory as a quantum mechanical system,''
Phys.\ Lett.\ B {\bf 558} (2003) 229
[arXiv:hep-th/0212269].

\bibitem{Roiban:2002xr}
R.~Roiban, M.~Spradlin and A.~Volovich,
``On light-cone SFT contact terms in a plane wave,''
JHEP {\bf 0310} (2003) 055
[arXiv:hep-th/0211220].

\bibitem{Gross:2002mh}
D.~J.~Gross, A.~Mikhailov and R.~Roiban,
``A calculation of the plane wave string Hamiltonian from N = 4 super-Yang-Mills theory,''
JHEP {\bf 0305} (2003) 025
[arXiv:hep-th/0208231].
%
T.~Klose,
``Conformal dimensions of two-derivative BMN operators,''
JHEP {\bf 0303}, 012 (2003)
[arXiv:hep-th/0301150].
%
J.~Gomis, S.~Moriyama and J.~Park,
``SYM description of pp-wave string interactions: Singlet sector and  arbitrary impurities,''
Nucl.\ Phys.\ B {\bf 665}, 49 (2003)
[arXiv:hep-th/0301250].
%
G.~Georgiou and V.~V.~Khoze,
``BMN operators with three scalar impurites and the vertex-correlator  duality in pp-wave,''
JHEP {\bf 0304}, 015 (2003)
[arXiv:hep-th/0302064].
%
D.~Z.~Freedman and U.~Gursoy,
``Instability and degeneracy in the BMN correspondence,''
JHEP {\bf 0308}, 027 (2003)
[arXiv:hep-th/0305016].
%
G.~Georgiou, V.~V.~Khoze and G.~Travaglini,
``New tests of the pp-wave correspondence,''
JHEP {\bf 0310}, 049 (2003)
[arXiv:hep-th/0306234].
%
P.~Bonderson,
``Decay modes of unstable strings in plane-wave string field theory,''
JHEP {\bf 0406}, 025 (2004)
[arXiv:hep-th/0307033].
%
P.~Gutjahr and J.~Plefka,
``Decay widths of three-impurity states in the BMN correspondence,''
arXiv:hep-th/0402211.
%
G.~Georgiou and G.~Travaglini,
``Fermion BMN operators, the dilatation operator of N = 4 SYM, and pp-wave string interactions,''
JHEP {\bf 0404}, 001 (2004)
[arXiv:hep-th/0403188].

\bibitem{Pankiewicz:2003pg}
A.~Pankiewicz,
``Strings in plane wave backgrounds,''
Fortsch.\ Phys.\  {\bf 51} (2003) 1139
[arXiv:hep-th/0307027].
%
J.~C.~Plefka,
``Lectures on the plane-wave string / gauge theory duality,''
Fortsch.\ Phys.\  {\bf 52} (2004) 264
[arXiv:hep-th/0307101].
%
J.~M.~Maldacena,
``TASI 2003 lectures on AdS/CFT,''
arXiv:hep-th/0309246.
%
M.~Spradlin and A.~Volovich,
``Light-cone string field theory in a plane wave,''
arXiv:hep-th/0310033.
%
D.~Sadri and M.~M.~Sheikh-Jabbari,
``The plane-wave / super Yang-Mills duality,''
arXiv:hep-th/0310119.
%
R.~Russo and A.~Tanzini,
``The duality between IIB string theory on pp-wave and N = 4 SYM: A status report,''
Class.\ Quant.\ Grav.\  {\bf 21} (2004) S1265
[arXiv:hep-th/0401155].

\bibitem{Callan:2003xr}
C.~G.~.~Callan, H.~K.~Lee, T.~McLoughlin, J.~H.~Schwarz, I.~Swanson and X.~Wu,
``Quantizing string theory in AdS(5) x S**5: Beyond the pp-wave,''
Nucl.\ Phys.\ B {\bf 673} (2003) 3
[arXiv:hep-th/0307032].
%
C.~G.~.~Callan, T.~McLoughlin and I.~Swanson,
``Holography beyond the Penrose limit,''
arXiv:hep-th/0404007.
%
C.~G.~.~Callan, T.~McLoughlin and I.~Swanson,
``Higher impurity AdS/CFT correspondence in the near-BMN limit,''
arXiv:hep-th/0405153.
%
I.~Swanson,
``On the integrability of string theory in AdS(5) x S**5,''
arXiv:hep-th/0405172.

\bibitem{Beisert:2002tn}
N.~Beisert,
``BMN operators and superconformal symmetry,''
Nucl.\ Phys.\ B {\bf 659} (2003) 79
[arXiv:hep-th/0211032].

\bibitem{Beisert:2003ys}
N.~Beisert,
``The su(2$|$3) dynamic spin chain,''
Nucl.\ Phys.\ B {\bf 682} (2004) 487
[arXiv:hep-th/0310252].

\bibitem{Gubser:2002tv}
S.~S.~Gubser, I.~R.~Klebanov and A.~M.~Polyakov,
``A semi-classical limit of the gauge/string correspondence,''
Nucl.\ Phys.\ B {\bf 636}, 99 (2002)
[arXiv:hep-th/0204051].
%
S.~Frolov and A.~A.~Tseytlin,
``Multi-spin string solutions in AdS(5) x S**5,''
Nucl.\ Phys.\ B {\bf 668}, 77 (2003)
[arXiv:hep-th/0304255].
%
S.~Frolov and A.~A.~Tseytlin,
``Semiclassical quantization of rotating superstring in AdS(5) x S(5),''
JHEP {\bf 0206}, 007 (2002)
[arXiv:hep-th/0204226].
%
G.~Arutyunov, S.~Frolov, J.~Russo and A.~A.~Tseytlin,
``Spinning strings in AdS(5) x S**5 and integrable systems,''
Nucl.\ Phys.\ B {\bf 671} (2003) 3
[arXiv:hep-th/0307191].

\bibitem{Minahan:2002ve}
J.~A.~Minahan and K.~Zarembo,
``The Bethe-ansatz for N = 4 super Yang-Mills,''
JHEP {\bf 0303}, 013 (2003)
[arXiv:hep-th/0212208].
%
N.~Beisert,
``The complete one-loop dilatation operator of N = 4 super Yang-Mills theory,''
Nucl.\ Phys.\ B {\bf 676}, 3 (2004)
[arXiv:hep-th/0307015].
%
N.~Beisert and M.~Staudacher,
``The N = 4 SYM integrable super spin chain,''
Nucl.\ Phys.\ B {\bf 670}, 439 (2003)
[arXiv:hep-th/0307042].
%
N.~Beisert,
``Higher loops, integrability and the near BMN limit,''
JHEP {\bf 0309}, 062 (2003)
[arXiv:hep-th/0308074].

\bibitem{Tseytlin:2003ii}
A.~A.~Tseytlin,
``Spinning strings and AdS/CFT duality,''
arXiv:hep-th/0311139.


\bibitem{Arutyunov:2003rg}
G.~Arutyunov and M.~Staudacher,
``Matching higher conserved charges for strings and spins,''
JHEP {\bf 0403}, 004 (2004)
[arXiv:hep-th/0310182].
%
M.~Kruczenski,
``Spin chains and string theory,''
arXiv:hep-th/0311203.
%
G.~Arutyunov and M.~Staudacher,
``Two-loop commuting charges and the string / gauge duality,''
arXiv:hep-th/0403077.

\bibitem{Kazakov:2004qf}
V.~A.~Kazakov, A.~Marshakov, J.~A.~Minahan and K.~Zarembo,
``Classical / quantum integrability in AdS/CFT,''
JHEP {\bf 0405}, 024 (2004)
[arXiv:hep-th/0402207].
%
M.~Kruczenski, A.~V.~Ryzhov and A.~A.~Tseytlin,
``Large spin limit of AdS(5) x S**5 string theory and low energy expansion of ferromagnetic spin chains,''
arXiv:hep-th/0403120.

\bibitem{Serban:2004jf}
D.~Serban and M.~Staudacher,
``Planar N = 4 gauge theory and the Inozemtsev long range spin chain,''
JHEP {\bf 0406}, 001 (2004)
[arXiv:hep-th/0401057].

\bibitem{Klose:2003qc}
T.~Klose and J.~Plefka,
``On the integrability of large N plane-wave matrix theory,''
Nucl.\ Phys.\ B {\bf 679} (2004) 127
[arXiv:hep-th/0310232].

\bibitem{Spradlin:2003bw}
M.~Spradlin and A.~Volovich,
``Note on plane wave quantum mechanics,''
Phys.\ Lett.\ B {\bf 565} (2003) 253
[arXiv:hep-th/0303220].

\bibitem{Metsaev:2002re}
R.~R.~Metsaev and A.~A.~Tseytlin,
``Exactly solvable model of superstring in plane wave Ramond-Ramond background,''
Phys.\ Rev.\ D {\bf 65}, 126004 (2002)
[arXiv:hep-th/0202109].

\bibitem{Mandelstam:hk}
S.~Mandelstam,
``Interacting String Picture Of The Neveu-Schwarz-Ramond Model,''
Nucl.\ Phys.\ B {\bf 69}, 77 (1974).
%
M.~B.~Green and J.~H.~Schwarz,
``Superstring Interactions,''
Nucl.\ Phys.\ B {\bf 218}, 43 (1983).

\bibitem{Cremmer:1974jq}
E.~Cremmer and .~L.~Gervais,
``Combining And Splitting Relativistic Strings,''
Nucl.\ Phys.\ B {\bf 76}, 209 (1974).
%
E.~Cremmer and .~L.~Gervais,
``Infinite Component Field Theory Of Interacting Relativistic Strings And Dual Theory,''
Nucl.\ Phys.\ B {\bf 90}, 410 (1975).

\bibitem{He:2002zu}
Y.~H.~He, J.~H.~Schwarz, M.~Spradlin and A.~Volovich,
``Explicit formulas for Neumann coefficients in the plane-wave geometry,''
Phys.\ Rev.\ D {\bf 67} (2003) 086005
[arXiv:hep-th/0211198].
%
J.~Lucietti, S.~Schafer-Nameki and A.~Sinha,
``On the plane-wave cubic vertex,''
arXiv:hep-th/0402185.

\bibitem{Greensite:1986gv}
J.~Greensite and F.~R.~Klinkhamer,
``New Interactions For Superstrings,''
Nucl.\ Phys.\ B {\bf 281} (1987) 269.
%
J.~Greensite and F.~R.~Klinkhamer,
``Contact Interactions In Closed Superstring Field Theory,''
Nucl.\ Phys.\ B {\bf 291} (1987) 557.
%
M.~B.~Green and N.~Seiberg,
``Contact Interactions In Superstring Theory,''
Nucl.\ Phys.\ B {\bf 299} (1988) 559.
%
J.~Greensite and F.~R.~Klinkhamer,
``Superstring Amplitudes And Contact Interactions,''
Nucl.\ Phys.\ B {\bf 304} (1988) 108.


\bibitem{Klebanov:2002mp}
I.~R.~Klebanov, M.~Spradlin and A.~Volovich,
``New effects in gauge theory from pp-wave superstrings,''
Phys.\ Lett.\ B {\bf 548} (2002) 111
[arXiv:hep-th/0206221].

\end{thebibliography}
\end{document}